\documentclass[pre,twocolumn,showpacs,floatfix,amsmath,amssymb]{revtex4}
\usepackage[latin1]{inputenc}
\usepackage[T1]{fontenc}
\usepackage[english]{babel}
\usepackage{graphicx}
\usepackage{psfrag}
\usepackage{amssymb}
\usepackage{amsmath}
\usepackage{amscd}
\usepackage{eucal}
\usepackage{color}
\usepackage{bm}
\newcommand{\bi}{\bibitem}

\begin{document}
\title{Mesoscopic Fluctuations of the Loschmidt Echo}
\author{Cyril Petitjean}
\author{Philippe Jacquod}
\affiliation{D\'epartement de Physique Th\'eorique,
Universit\'e de Gen\`eve, CH-1211 Gen\`eve 4, Switzerland}
\date{October 13 2004}
\begin{abstract}
We investigate the time-dependent variance of the fidelity 
with which an initial narrow wavepacket is reconstructed after
its dynamics is time-reversed with a perturbed Hamiltonian. In
the semiclassical regime of perturbation, we show that
the variance first rises algebraically up to a critical time $t_c$, after which
it decays. To leading order in the effective
Planck's constant $\hbar_{\rm eff}$, this decay is given by the sum of 
a classical term $\simeq \exp[-2 \lambda t]$, a
quantum term $\simeq 2 \hbar_{\rm eff} \exp[-\Gamma t]$
and a mixed term $\simeq 2 \exp[-(\Gamma+\lambda)t]$.
Compared to the behavior of the average fidelity, this allows 
for the extraction of the classical Lyapunov exponent $\lambda$
in a larger parameter range. Our results are confirmed by numerical
simulations.
\end{abstract}
\maketitle{}
Fluctuations of a physical quantity often contain more information 
than its average. For example, quantum signatures of classical
chaos are absent of the average density of states, but strongly 
affect spectral fluctuations \cite{Haake00}. 
In the search for such signatures, 
another approach has been to investigate
the sensitivity to
an external perturbation that is exhibited by the quantum dynamics
\cite{schack}. Going back to Ref.~\cite{peres}, the central
quantity in this approach is
the Loschmidt Echo \cite{Jal01}, the fidelity
\begin{equation}\label{fidelity}
M(t) = |\langle \psi_0 | \exp[i Ht] \exp[-iH_0t]|\psi_0\rangle|^2
\end{equation}
with which an initial quantum state $\psi_0$ is reconstructed after
the dynamics is time-reversed using a perturbed Hamiltonian,
$H=H_0 + \epsilon V$ (we set $\hbar \equiv 1$). 
This approach proved very fruitful, however,
most investigations of $M(t)$ (which we will briefly summarize
below) considered the properties of the
{\em average} fidelity $\overline{M(t)}$,
either over different $\psi_0$, or different elements
of an ensemble of unperturbed Hamiltonians $H_0$ 
(having for instance the same classical Lyapunov  
exponent $\lambda$) and/or perturbation
$V$. Curiously enough, the variance $\sigma^2(M)$ 
of the fidelity has been largely
neglected so far. The purpose of this
article is to fill this gap. We will see that the 
variance $\sigma^2(M)$ has a much richer behavior than $\overline{M(t)}$,
allowing for the
extraction of $\lambda$ in a larger parameter range, and exhibiting
a nonmonotonous behavior with a non-self-averaging maximal value
$\sigma(t_c)/\overline{M(t_c)} \simeq 1$.

We first summarize what is known about the average fidelity 
$ \overline{M (t)}$ in quantum 
chaotic systems. Three regimes of
perturbation strength are differentiated by three energy scales \cite{Jac01}:
the energy bandwidth $B$ of $H_0$,
the golden rule spreading $\Gamma= 2 \pi \epsilon^2 \overline{|\langle
\phi^{(0)}_\alpha|V|\phi^{(0)}_\beta \rangle|^2}/\Delta$
of an eigenstate $\phi^{(0)}_{\alpha}$  of $H_0$ over the eigenbasis
$\{\phi_{\alpha}\}$
of $H$, and the level spacing $\Delta = B \hbar_{\rm eff}$ 
($\hbar_{\rm eff}=\nu^d/\Omega$
is the effective Planck's constant, given by the ratio
of the wavelength volume to the
system's volume). These three regimes are
(i) the weak perturbation regime $\Gamma < \Delta$, with a typical 
Gaussian decay 
$ \overline{M(t)} \simeq \exp(-\overline{\Sigma^2} t^2)$,
$\Sigma^2 \equiv \epsilon^2 (\langle\psi_0|V^2|\psi_0\rangle
-\langle\psi_0|V|\psi_0\rangle^2) $,
$\overline{\Sigma^2}\simeq \Gamma \Delta \hbar_{\rm eff}^{-1}$ 
\cite{peres,Seligman02} (corrections to 
this Gaussian decay have been discussed in Ref.~\cite{steve_semicl}),
(ii) the semiclassical golden rule regime $\Delta < \Gamma < B$, where
the decay is exponential with a rate set by the smallest
of $\Gamma$ and $\lambda$, $ \overline{M(t)} 
\simeq \exp[-{\rm min}(\Gamma,\lambda) t]$
\cite{Jal01,Jac01,cucchietti}, and (iii) the strong perturbation regime
$\Gamma > B$ with another Gaussian decay  
$ \overline{M(t)} \simeq \exp(-B^2 t^2)$
\cite{Jac01}. This classification
is based on the scheme of Ref.~\cite{Jac01} which relates the 
behavior of $\overline{M(t)}$ to the local spectral density of 
eigenstates of $H_0$ over the eigenbasis of $H$ \cite{Jac01,Wis01}.
Accordingly, regime (ii) corresponds to the range
of validity of Fermi's golden rule, 
where the local spectral density has a Lorentzian
shape \cite{wigner,Jac01,Wis01}.  Quantum disordered systems 
with diffractive impurities, on the other hand,
have been predicted to exhibit golden rule decay $\propto
\exp[-\Gamma t]$ and Lyapunov decay $\propto \exp[-\lambda t]$ 
in different time intervals for a single set of 
parameters \cite{adamov}. It is also worth mentioning that
regular systems exhibit a very different behavior,
where in the semiclassical regime (ii), $\overline{M(t)}$ 
decays as a power-law \cite{Inanc03} (see also Ref.~\cite{emerson}).
Finally, while in chaotic systems the averaging procedure has been
found to be ergodic, i.e. considering different states $\psi_0$
is equivalent to considering different realizations of $H_0$ or
$V$, the Lyapunov decay exists only for specific choices where
$\psi_0$ has a well defined classical
meaning, like a coherent or a position state
\cite{Jal01,jiri,Inanc02,iomin}.

Investigations beyond this qualitative picture 
have focused on crossover regions between the regimes (i) and 
(ii) \cite{steve_semicl} and deviations from the behavior (ii)
$\simeq \exp[-{\rm min}(\Gamma,\lambda)t]$ due to action 
correlations in weakly chaotic systems \cite{casati}.
Ref.~\cite{silvestrov} provides the only analytical 
investigation of fluctuations of $M(t)$ to date. It shows
that, for classically large perturbations, $\Gamma \gg B$, 
$\overline{M(t)}$ is dominated
by very few exceptional events, so that a typical $\psi_0$'s
fidelity is better described by 
$\exp[\overline{\ln(M)}]$, and that $M(t)$ does not
fluctuate after the Ehrenfest time
$t_E = \lambda^{-1} |\ln[\hbar_{\rm eff}]|$. We will
see that these conclusions do not apply to the regime (ii) of present
interest. While some numerical data 
for the distribution of $M(t)$ in the
weak perturbation regime (i) were presented in Ref.~\cite{gorin},
we focus here on chaotic
systems and investigate the behavior of $\sigma^2(M)$ in the 
semiclassical regime (ii). 

We first follow a semiclassical approach along the lines of 
Ref.~\cite{Jal01}. We consider an initial
Gaussian wavepacket $\psi_0({\bf r}_0') = 
(\pi \nu^2)^{-d/4} \exp[i {\bf p}_0 \cdot ({\bf r}_0'-{\bf r}_0)-
|{\bf r}_0'-{\bf r}_0|^2/2 \nu^2]$, and approximate its
time-evolution by 
\begin{subequations}
\begin{eqnarray}\label{propwp}
\langle {\bf r}|
\exp(-i H_0 t) |\psi_0\rangle  =  \int d{\bf r}_0'
\sum_s K_s^{H_0}({\bf r},{\bf r}_0';t) \psi_0({\bf r}_0'), \nonumber \\
K_s^{H_0}({\bf r},{\bf r}_0';t)  =  \frac{C_s^{1/2}}{(2 \pi i)^{d/2}} 
\exp[i S_s^{H_0}({\bf r},{\bf r}_0';t)-i \pi \mu_s/2].\nonumber 
\end{eqnarray}
\end{subequations}
\noindent The semiclassical propagator is expressed
as a sum over classical trajectories (labelled $s$)
connecting ${\bf r}$ and ${\bf r}_0'$ in the time $t$.
For each $s$, the partial propagator contains
the action integral $S_s^H({\bf r},{\bf r}_0';t)$ along $s$,
a Maslov index $\mu_s$, and
the determinant $C_s$ of the stability matrix \cite{chaosbook}.
We recall that this approach allows to calculate the time
evolution of smooth, localized wavepackets 
up to algebraically long times
$\propto {\cal O}(\hbar_{\rm eff}^{-a}) \gg t_E$
(with $a>0$)\cite{steve}.

The fidelity then reads,
\begin{eqnarray}\label{moft}
M(t) & = & \Bigg|\int d{\bf r}_1 \int d{\bf r}_0' \int d{\bf r}_0'' \;
\psi_0({\bf r}_0')\psi_0^*({\bf r}_0'')  \\
&& \times \sum_{s_1,s_2} K_{s_1}^{H_0}({\bf r}_1,{\bf r}_0 ';t) \;
 [K_{s_2}^H({\bf r}_1,{\bf r}_0'';t)]^* \Bigg|^2 \nonumber 
\end{eqnarray}
We want to calculate $M^2(t)$. Squaring Eq.~(\ref{moft}), we see
that $M^2(t)$ is given by
eight sums over classical paths and twelve spatial integrations.
Noting that $\psi_0$ is
a narrow Gaussian wavepacket, we first linearize
all eight action integrals around ${\bf r}_0$, 
\begin{equation}
S_{s}({\bf r},{\bf r}_0';t) \simeq S_{s}({\bf r},{\bf r}_0;t)
-({\bf r}_0'-{\bf r}_0) \cdot {\bf p}_{s}.
\end{equation}
We can then perform the Gaussian integrations over the eight initial positions
${\bf r}_0'$, ${\bf r}_0''$ and so forth. In this way 
$M^2(t)$ is expressed as a sum over eight trajectories connecting
${\bf r}_0$ to four
independent final points ${\bf r}_j$ over which one integrates,
\begin{eqnarray}\label{eq:m2}
M^2(t) &=& \int \prod_{j=1}^4 d{\bf r}_j
\sum_{s_i;i=1}^8 \;  \exp[i (\Phi^{H_0}-\Phi^{H}-\pi {\mathcal M}/2)] 
\nonumber \\
&& \times  \left(\prod_i C_{s_i}^{1/2} \left(\frac{\nu^2}{\pi}\right)^{d/4}
\exp(-\nu^2\delta{\bf p}_{s_i}^2/2)\right),
\end{eqnarray}
where we introduced ${\mathcal M}=\sum_{i=0}^{3} (-1)^i (\mu_{s_{2i+1}} -
\mu_{s_{2i+2}} )$ and $\delta{\bf p}_{s_i}={\bf p}_{s_i}-{\bf p}_0$.

\begin{figure}
\begin{psfrags}
\psfrag{ro}{${\bf r}_0$}
\psfrag{r1}{${\bf r}_1$}
\psfrag{r2}{${\bf r}_2$} 
\psfrag{r3}{${\bf r}_3$}
\psfrag{r4}{${\bf r}_4$}
\psfrag{s1}{$ s_1 $}
\psfrag{s2}{$ s_2 $} 
\psfrag{s3}{$ s_3 $}
\psfrag{s4}{$ s_4 $}
\psfrag{s5}{$ s_5 $}
\psfrag{s6}{$ s_6 $} 
\psfrag{s7}{$ s_7 $}
\psfrag{s8}{$ s_8 $}
\includegraphics[height=5cm]{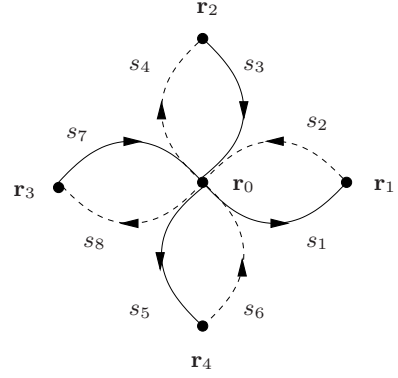}
\end{psfrags}
\caption{\label{fig1} Diagrammatic representation of the 
squared fidelity 
$M^2(t)$.}
\end{figure}
 
\begin{figure*}
\begin{center}\nonumber 
\begin{psfrags}
\psfrag{ro}{${\bf r}_0$}
\psfrag{r1}{${\bf r}_1$}
\psfrag{r2}{${\bf r}_2$} 
\psfrag{r3}{${\bf r}_3$}
\psfrag{r4}{${\bf r}_4$} 
\psfrag{r13}{${\bf r}_1\! \simeq \!{\bf r}_3$}
\psfrag{r12}{${\bf r}_1\! \simeq\!{\bf r}_2$}
\psfrag{r14}{${\bf r}_1\! \simeq\!{\bf r}_4$}  
\psfrag{r24}{${\bf r}_2\! \simeq \!{\bf r}_4$}
\psfrag{SIGMA}{$\sigma^2(M)$}
\psfrag{LYAPDOUB}{$ { \alpha^2 e^{-2\lambda t} }$}
\psfrag{LYAPSINGLET}{${ 2 \alpha e^{-\lambda t}e^{-\Gamma t } }$}
\psfrag{SATSINGLET}{${ 2\hbar_{\rm eff}e^{-\Gamma t }\Theta(t-t_E) }$}
\psfrag{SATSAT}{${ \hbar_{\rm eff}^2\Theta(t-t_E) }$}
\includegraphics[height=3.cm]{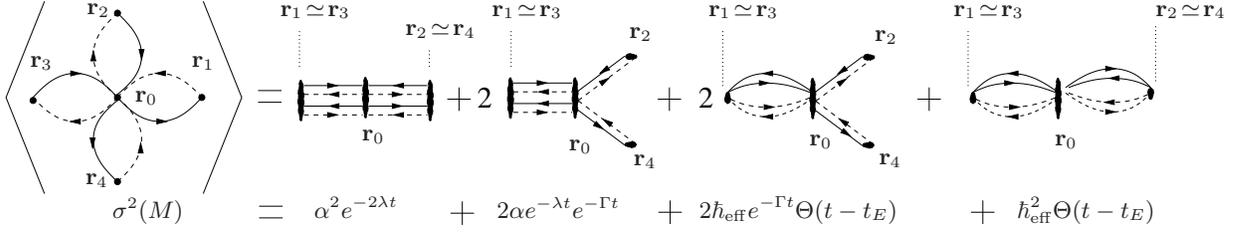}
\end{psfrags}
\end{center}
\vspace{-0.4cm}
\caption{\label{fig2} Diagrammatic representation of the 
averaged fidelity variance
$\sigma^2(M)$ and the three time-dependent contributions that 
dominate semiclassically, together with the contribution giving the
long-time saturation of $\sigma^2(M)$.}
\end{figure*}

The expression of Eq.~(\ref{eq:m2}) 
is schematically described in Fig.~\ref{fig1}. 
Classical trajectories are represented
by a full line if they correspond to $H_0$
and a dashed line for $H$, with an arrow indicating the direction
of propagation. 
In the semiclassical limit $S_s \gg 1$ (we recall that actions are expressed
in units of $\hbar$), Eq.~(\ref{eq:m2}) is dominated by terms which satisfy
a stationary phase
condition, i.e. where the variation of the difference of the two action phases
\begin{subequations}
\begin{eqnarray}
\Phi^{H_0}&=&S^{H_0}_{s_1}({\bf r}_1,{\bf r}_0;t)-
S^{H_0}_{s_3}({\bf r}_0,{\bf r}_2;t) \nonumber\\ 
&+&
S^{H_0}_{s_5}({\bf r}_4,{\bf r}_0;t)-
S^{H_0}_{s_7}({\bf r}_0,{\bf r}_3;t), \\
\Phi^{H}&=&
S^{H}_{s_2}({\bf r}_0,{\bf r}_1;t)\;-
S^{H}_{s_4}({\bf r}_2,{\bf r}_0;t)\;\nonumber\\ 
&+&
S^{H}_{s_6}({\bf r}_0,{\bf r}_4;t)\;-
S^{H}_{s_8}({\bf r}_3,{\bf r}_0;t),
\end{eqnarray}
\end{subequations}
has to be minimized. These stationary phase
terms are easily identified from the diagrammatic
representation as those where two classical trajectories $s$ and $s'$ 
of opposite direction of propagation
are {\em contracted}, i.e. $s=s'$, up to a quantum resolution given
by the wavelength $\nu$ \cite{caveat2}. This is represented in
Fig.~\ref{fig2} by bringing two lines together in parallel.
Contracting either two dashed or two full lines allows for
an almost exact cancellation of the actions, hence an almost 
perturbation-independent contribution, up to a contribution arising
from the finite resolution $\nu$ with which the two paths overlap. 
However when a full line is
contracted with a dashed line, the resulting contribution still
depends on the action $\delta S_s = -\epsilon \int_s V({\bf q}(t),t)$
accumulated by the perturbation along the classical path $s$,
spatially parametrized as ${\bf q}(t)$.
Since we are interested in the variance $\sigma^2(M) = 
\overline{M^2}-\overline{M}^2$ (this is indicated by brackets
in Fig.~\ref{fig2}) we must subtract the
terms contained in $\overline{M}^2$
corresponding to independent contractions in each of the two
subsets $(s_1,s_2,s_3,s_4)$ and $(s_5,s_6,s_7,s_8)$. Consequently,
all contributions to $\sigma^2(M)$ require pairing of spatial
coordinates, $|{\bf r}_i-{\bf r}_j| \le \nu$, for at least one
pair of indices $i,j=1,2,3,4$.

With these considerations, the four dominant contributions to 
$\sigma^2(M)$ are depicted on the right-hand side of Fig.~\ref{fig2}.
The first one corresponds to $s_1=s_2 \simeq
s_7=s_8$ and $s_3=s_4 \simeq s_5=s_6$, which
requires ${\bf r}_1 \simeq {\bf r}_3$, ${\bf r}_2 \simeq {\bf r}_4$.
This gives a contribution
\begin{eqnarray}\label{sigma1}
\sigma^2_1 & = & \left(\frac{\nu^2}{\pi}\right)^{2d} \Bigg\langle 
\int d{\bf r}_1 d{\bf r}_3 \sum
C^2_{s_1} \\
& & \times
\exp[-2 \nu^2 \delta{\bf p}_{s_1}^2 + i \delta \Phi_{s_1}]
\Theta(\nu-|{\bf r}_1-{\bf r}_3|)
\Bigg\rangle^2, \nonumber
\end{eqnarray}
where $\delta \Phi_{s_1}= \epsilon \int_0^t dt' \nabla V[{\bf q}(t')] [
{\bf q}_{s_1}(t')-{\bf q}_{s_7}(t')]$ arises from the linearization of $V$
on $s=s_{1,2} \simeq s'=s_{7,8}$ \cite{Jal01,jiri}, and
${\bf q}_{s_1}(\tilde{t})$ lies on $s_1$ with ${\bf q}(0)={\bf r}_0$
and ${\bf q}(t)={\bf r}_1$. In Eq.~(\ref{sigma1})
the integrations are restricted by $|{\bf r}_1-{\bf r}_3|\le \nu$ because
of the finite resolution with which two paths can be equated (this
is also enforced by the presence of $\delta \Phi_s$ as we will see
momentarily).
For long enough times, $t \gg t^*$, 
the phases $\delta \Phi_s$ fluctuate randomly and exhibit no correlation 
between different trajectories \cite{caveat0}. 
One thus applies the Central Limit Theorem
(CLT) $\langle \exp[i \delta \Phi_s] \rangle = \exp[-\langle \delta \Phi_s^2 
\rangle/2] \simeq \exp[-\epsilon^2 \int dt \langle \nabla V(0) \cdot 
\nabla V(t) \rangle |{\bf r}_1-
{\bf  r}_3|^2/2 \lambda]$. After performing a change 
of integration variable $\int d{\bf r} \sum_s C_s = \int d{\bf p}$ and
using the asymptotic expression
$C_s \simeq (m/t)^{d} \exp[-\lambda t]$ \cite{chaosbook}, one gets
\begin{subequations}\label{sigma1lambda}
\begin{eqnarray}
\sigma_1^2&=& \alpha^2 \exp[-2\lambda t], \\
\alpha &=& \left( { \lambda \nu^2 m^2 \over \epsilon^2 t^2 \int 
d\tau \langle \nabla V(0) 
\cdot \nabla V(\tau) \rangle }\right)^{d/2}.
\end{eqnarray}
\end{subequations}

The second dominant term is obtained from  
$s_1=s_2 \simeq s_7=s_8$, $s_3=s_4$ and $s_5=s_6$, with
${\bf r}_1 \simeq {\bf r}_3$, or equivalently $s_1=s_2$, $s_7=s_8$ and  
$s_3=s_4 \simeq s_5=s_6$ with ${\bf r}_2 \simeq {\bf r}_4$. Therefore
this term comes with a multiplicity of two, and one obtains
\begin{eqnarray}\label{sigma2}
\sigma^2_2 &=&  2 \left(\frac{\nu^2}{\pi}\right)^{2d} 
\Bigg\langle
\int d{\bf r}_1 d{\bf r}_3 \sum C^2_{s_1} \nonumber \\
&& \times \exp[-2 \nu^2 \delta{\bf p}_{s_1}^2
+i\delta \Phi_{s_1}] \Theta(\nu -|{\bf r}_1-{\bf r}_3|)
\Bigg\rangle \nonumber \\
&& \times \left\langle\int d{\bf r}_2 \sum C_{s_3} 
\exp[-\nu^2 \delta{\bf p}_{s_3}^2+i \delta S_{s_3}] 
\right\rangle^2,
\end{eqnarray}
again with the restriction $|{\bf r}_1-{\bf r}_3|\le \nu$.
To calculate the first bracket on the right-hand side of
Eq.~(\ref{sigma2}), we first average the complex exponential, assuming 
again that enough time has elapsed so that actions are randomized. 
The CLT gives $\langle \exp[i \delta S_{s_3}] \rangle = 
\exp(-{\textstyle \frac{1}{2}} \langle \delta S_{s_3}^2 \rangle )$ with
\begin{equation}\label{phase}
\langle \delta S_{s_3}^2 \rangle  = \mbox{}
\epsilon^2
\int_0^t d{\tilde t} \int_0^t d{\tilde t}' 
\langle V[{\bf q}(\tilde{t})]
V[{\bf q}(\tilde{t}')] \rangle.
\end{equation}
Here ${\bf q}(\tilde{t})$ lies on $s_3$ with ${\bf q}(0)={\bf r}_0$
and ${\bf q}(t)={\bf r}_2$. In hyperbolic systems,
correlators typically decay exponentially fast,
\begin{equation}
\langle V[{\bf q}(\tilde{t})]
V[{\bf q}(\tilde{t}')] \rangle \propto \exp[-\eta |t-t'|],
\end{equation} 
with an upper bound on $\eta$ set by the smallest positive Lyapunov
exponent \cite{eckmann}. One thus obtains
$\langle \delta S_{s_3}^2 \rangle = \Gamma t$.
Usually $\Gamma \propto \epsilon^2$ is
identified with the golden rule spreading of
eigenstates of $H$ over those of $H_0$ \cite{steve_semicl,Jac01}.
It is dominated by the short-time behavior of
$\langle V[{\bf q}(\tilde{t})] V[{\bf q}(0)] \rangle$.
We stress however that for long enough times,
$\langle \delta S_{s_3}^2 \rangle \propto t$
still holds to leading order even with a power-law decay of the
correlator $ \langle V[{\bf q}(\tilde{t})]
V[{\bf q}(\tilde{t}')] \rangle \propto |t-t'|^{-\eta}$,
provided $\eta$ is sufficiently large, $\eta \ge 1$. 
We note that similar expressions as Eq.~(\ref{phase})
relating the decay of $\overline{M}$ to
time integrations over the perturbation correlator have been
derived in Refs.~\cite{Seligman02,gorin} using a different approach than
the semiclassical method of Ref.~\cite{Jal01} used here.
Further using the sum rule 
\begin{equation}\label{eq:sumrule}
(\nu^2/\pi)^{d} \left(\int d{\bf r} \sum C_{s} 
\exp[-\nu^2 \delta{\bf p}_{s}^2 ] \right)^2=1, 
\end{equation}
one finally obtains
\begin{equation}\label{sigma2lg}
\sigma_2^2 = 2 \alpha \exp[-\lambda t] \exp[-\Gamma t].
\end{equation}

The third and last dominant time-dependent term arises
from either $s_1=s_7$, $s_2=s_8$, $s_3=s_4$, $s_5=s_6$ and 
${\bf r}_1\simeq{\bf r}_3$, or $s_1=s_2$, $s_3=s_5$, $s_4=s_6$, $s_7=s_8$ and 
${\bf r}_2\simeq{\bf r}_4$. It thus also has a multiplicity of two and reads
\begin{eqnarray}\label{sigma3}
\sigma^2_3 &=& 2 \left(\frac{\nu^2}{\pi}\right)^{2d}
\Big\langle
\int d{\bf r}_1 d{\bf r}_2 d{\bf r}_3 d{\bf r}_4
\sum C_{s_1}C_{s_2}C_{s_3}C_{s_5}  \nonumber \\
&& \times \exp[-\nu^2 (\delta{\bf p}_{s_1}^2+
\delta{\bf p}_{s_2}^2+\delta{\bf p}_{s_3}^2+\delta{\bf p}_{s_5}^2)] 
\nonumber \\
&& \times \exp[i (\delta S_{s_3}-\delta S_{s_5})] \;
\Theta(\nu - |{\bf r}_1-{\bf r}_3|) \Big\rangle.
\end{eqnarray}
The integrations, again, have to be performed with
$|{\bf r}_1-{\bf r}_3|\le \nu$. We incorporate this restriction
in the calculation by making the ergodicity assumption, setting
\begin{eqnarray}\label{saturation}
&&\langle \int d{\bf r}_1 d{\bf r}_2 d{\bf r}_3 d{\bf r}_4 \ldots
\Theta(\nu - |{\bf r}_1-{\bf r}_3|) \rangle
\nonumber \\
&=& \hbar_{\rm eff} \langle \int d{\bf r}_1 d{\bf r}_2 d{\bf r}_3 
d{\bf r}_4 \ldots \rangle \Theta(t-t_E), 
\end{eqnarray}
which is valid for
times larger than the Ehrenfest time \cite{Berman78} (for shorter times,
$t<t_E$, the third diagram on the right-hand side of Fig.~\ref{fig2} 
goes into the second one). One then 
averages the phases using the CLT to get
\begin{eqnarray}\label{sigma3g}
\sigma^2_3 = 2 \hbar_{\rm eff} \exp[-\Gamma t]
\Theta(t-t_E).
\end{eqnarray}

Subdominant terms are obtained by
higher-order contractions (e.g. setting ${\bf r}_2 \simeq {\bf r}_4$
in the second and third graphs on the right hand-side
of Fig.\ref{fig2}). They either decay faster, or are of higher
order in $\hbar_{\rm eff}$, or both. We only discuss the
term which gives the long-time saturation at the ergodic value
$\sigma^2(M) \simeq \hbar_{\rm eff}^{2}$. For 
$t>t_E$, there is a phase-free (and hence time-independent)
contribution with four different paths, resulting
from the contraction $s_1=s_7$, $s_2=s_8$, $s_3=s_5$, $s_4=s_6$,
and ${\bf r}_1 \simeq {\bf r}_3$, ${\bf r}_2 \simeq {\bf r}_4$.
Its contribution is sketched as the fourth diagram on the
right-hand side of Fig.~\ref{fig2}. It gives
\begin{eqnarray}\label{eq:satsat} 
\sigma^2_4 &= &  \left(\frac{\nu^2}{\pi}\right)^{2d} 
\Big\langle\int d{\bf r}_1d{\bf r}_3 \sum C_{s_1}C_{s_2} \\
&& \times \exp[-\nu^2 (\delta{\bf p}_{s_1}^2  +\delta{\bf p}_{s_2}^2 )]  
\Theta(\nu-|{\bf r}_1-{\bf r}_3|)\Big\rangle^2. \nonumber
\end{eqnarray}
From the sum rule of Eq.~(\ref{eq:sumrule}), and again invoking the
long-time ergodicity of the semiclassical dynamics, Eq.~(\ref{saturation}), 
one obtains the long-time saturation of $\sigma^2(M)$,
\begin{equation}\label{eq:saturation}
\sigma^2_4 = \hbar_{\rm eff}^2 \Theta(t-t_E).
\end{equation}
Note that for $t<t_E$, this contribution does not exist by itself and is 
included in $\sigma_1^2$, Eq.~(\ref{sigma1lambda}).

According to our semiclassical approach,
the fidelity has a variance given to leading order by the sum of the
four terms of 
Eqs.~(\ref{sigma1lambda}), (\ref{sigma2lg}), (\ref{sigma3g}) and
(\ref{eq:saturation})
\begin{eqnarray}\label{sigmascl}
\sigma^2_{\rm sc} &=& \alpha^2 \exp[-2 \lambda t] + 
2 \alpha \exp[-(\lambda + \Gamma) t] \\
&& + 2 \hbar_{\rm eff} \exp[-\Gamma t] \Theta(t-t_E)
+\hbar_{\rm eff}^2 \Theta(t-t_E). \nonumber
\end{eqnarray}
Eq.~(\ref{sigmascl}) is the central result of this paper.
We see that 
for short enough times, i.e. before ergodicity and the saturation
of $M(t) \simeq \hbar_{\rm eff}$ and $\sigma^2(M) \simeq \hbar_{\rm eff}^{2}$
is reached, the first term on the right-hand
side of (\ref{sigmascl}) will dominate as long as $\lambda < \Gamma$.
For $\lambda > \Gamma$ on the other hand, $\sigma^2(M)$ exhibits
a behavior $\propto \exp[-(\lambda+\Gamma) t]$ for $t<t_E$, turning
into $\propto \hbar_{\rm eff} \exp[-\Gamma t]$ for $t>t_E$. 
Thus, contrary to $\overline{M}$, $\sigma^2(M)$ allows to
extract the Lyapunov exponent from the second term on the right-hand
side of Eq.~(\ref{sigmascl}) even when $\lambda > \Gamma$. Also
one sees that, unlike the strong perturbation regime $\Gamma \gg B$
\cite{silvestrov}, $M(t)$ continues
to fluctuate above the residual variance $\simeq \hbar_{\rm eff}^2$ up to
a time $ \simeq \Gamma^{-1} |\ln \hbar_{\rm eff}|$
in the semiclassical regime $B>\Gamma>\Delta$. For $\Gamma \ll \lambda$,
$\Gamma^{-1} |\ln \hbar_{\rm eff}| \gg t_E$ and  $M(t)$ fluctuates 
beyond $t_E$.

The above semiclassical approach breaks down at short
times for which not enough phase is accumulated to motivate
a stationary phase approximation \cite{caveat}. 
To get 
the short-time behavior of $\sigma^2(M)$, we instead
Taylor expand the time-evolution exponentials 
$\exp[\pm iH_{(0)}t]=1 \pm i H_{(0)}t - H_{(0)}^2 t^2 /2 + ...
+ O(H_{(0)}^5 t^5)$. The resulting expression 
for $\sigma^2(M)$ contains matrix elements such
as $\langle \psi_0 | H_{(0)}^a | \psi_0 \rangle$, $a=1,2,3,4$, which 
one then calculates using a Random Matrix Theory (RMT) approach \cite{mehta}
for the chaotic quantized Hamiltonian $H_{(0)}$ 
\cite{Jac01,cucchietti,gorin}.
Keeping non-vanishing terms of lowest order in $t$,
one has a quartic onset 
$\sigma^2(M) \simeq (\overline{\Sigma^4}-\overline{\Sigma^2}^2)t^4$ for
$t \ll \Sigma^{-1}$, with $\Sigma^a \equiv 
[\epsilon^2 (\langle\psi_0|V^2|\psi_0\rangle
-\langle\psi_0|V|\psi_0\rangle^2)]^{a/2} $. RMT gives
$(\overline{\Sigma^4}-\overline{\Sigma^2}^2) \propto (\Gamma B)^2$,
with a system-dependent prefactor of order one.
From this and Eq.~(\ref{sigmascl}) one concludes that
$\sigma^2(M)$ has a nonmonotonous behavior, i.e. it first rises
at short times, until it decays after a time $t_c$ which one can evaluate
by solving $\sigma^2_{\rm sc}(t_c) = (\Gamma B)^2 t_c^4$.
In the regime $B>\Gamma >\lambda$ one gets
\begin{eqnarray}\label{tc}
t_c&=&\left(\frac{\alpha_0}{\Gamma B}\right)^{1/2+d}
\Bigg[1-\lambda \left(\frac{\alpha_0}{\Gamma B}\right)^{1/2+d}\frac{1}{2+d}
\nonumber \\
&& +O\left(\lambda^2
\left\{\frac{\alpha_0}{\Gamma B}\right\}^{2/2+d} \right) \Bigg], 
\end{eqnarray}
and thus
\begin{eqnarray}\label{sigmatc}
\sigma^2(t_c) &\simeq& (\Gamma B)^2
\left(\frac{\alpha_0}{\Gamma B}\right)^{4/2+d} \Bigg[
1-\frac{4 \lambda}{2+d}  
\left(\frac{\alpha_0}{\Gamma B}\right)^{1/2+d} \nonumber \\
&& +O\left(\lambda^2
\left\{\frac{\alpha_0}{\Gamma B}\right\}^{2/2+d} \right)  \Bigg].
\end{eqnarray}
We explicitely took the $t$-dependence  
$\alpha(t) = \alpha_0 t^{-d}$ into account. We estimate that
$\alpha_0 \propto (\Gamma \lambda)^{-d/2}$ (obtained
by setting the Lyapunov time
equal to few times the time of flight through a correlation
length of the perturbation potential, as is the case for
billiards or maps), to get
$\sigma^2(t_c) \propto (B/\lambda)^{2d/2+d} \gg 1$. Because
$0 \le M(t) \le 1$, this value is
however bounded by $\overline{M}^2(t_c)$. Since in the other regime
$\Gamma \ll \lambda$, one has $\sigma^2(t_c) \simeq
2 \hbar_{\rm eff} [1-(2 \hbar_{\rm eff})^{1/4} \sqrt{\Gamma/B}]$
we predict that $\sigma^2(t_c)$ grows during the crossover from
$\Gamma \ll \lambda$ to $\Gamma > \lambda$, until it saturates at a
non-self-averaging value, $\sigma(t_c)/\overline{M}(t_c) \approx 1$,
independently on $\hbar_{\rm eff}$ and $B$, with possibly a weak
dependence on $\Gamma$ and $\lambda$. 

We conclude this analytical section by
mentioning that applying the RMT approach to longer times
reproduces Eq.~(\ref{sigmascl}) with $\lambda \rightarrow \infty$ 
\cite{cyril}. This reflects the fact that RMT is strictly recovered 
for $t_E = 0$ only. 

\begin{figure}
\includegraphics[width=6.2cm]{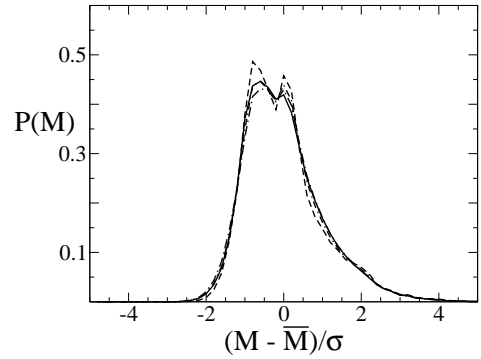}
\caption{\label{fig3}
Distribution $P(M)$ of the fidelity computed
for $10^4$ different $\psi_0$
for $N=32768$, $\delta K=5.75 \cdot 10^{-5}$ (i.e. $\Gamma \approx 0.09$),
at times $t=25,50,75$ and 100 kicks.\\[-1mm]}
\end{figure}

To illustrate our results, we present some numerical data.
We based our simulations on the kicked rotator model with Hamiltonian
\cite{felix}
\begin{equation}\label{kickrot}
H_0 = \frac{\hat{p}^2}{2} + K_0 \cos \hat{x} \sum_n \delta(t-n).
\end{equation}
We concentrate on the regime $K > 7$, for which the dynamics is fully
chaotic with a Lyapunov exponent $\lambda = \ln[K/2]$. 
We quantize this Hamiltonian on a torus, which 
requires to consider discrete values
$p_l=2 \pi l/N$ and $x_l=2 \pi l/N$, $l=1,...N$, hence $\hbar_{\rm eff}=
1/N$. The fidelity 
(\ref{fidelity}) is computed for discrete times $t=n$, as
\begin{eqnarray}\label{krot_fid}
M(n) & = & |\langle \psi_0 | \left(U_{\delta K}^*\right)^n 
\left(U_{0}\right)^n  | \psi_0 \rangle|^2
\end{eqnarray}
using
the unitary Floquet
operators $U_{0}=\exp[-i \hat{p}^2/2 \hbar_{\rm eff}] \exp[-iK_0 \cos \hat{x}/ 
\hbar_{\rm eff}]$ and $U_{\delta K}$ having
a perturbed Hamiltonian $H$ with $K=K_0 + \delta K$. 
The quantization procedure results in a matrix form of the Floquet operators,
whose matrix elements in $x-$representation are given by
\begin{eqnarray}
\left(U_{0}\right)_{l,l'} & = & \frac{1}{\sqrt{N}} \exp[i 
\frac{\pi (l-l')^2}{N}] \exp[-i \frac{N K_0}{2 \pi} \cos \frac{2 \pi l'}{N}].
\nonumber
\end{eqnarray}
The local spectral density of eigenstates of $U_{\delta K}$
over those of $U_0$ has a Lorentzian shape with a width $\Gamma \propto (\delta
K/\hbar_{\rm eff})^2$ (there is a weak dependence of $\Gamma$ in $K_0$) in the 
range $B = 2 \pi \gtrsim  \Gamma >\Delta = 2 \pi/N$). This is illustrated
in the inset to Fig.~\ref{fig6}. 

Numerically, the time-evolution of $\psi_0$ in
the fidelity, Eq.~(\ref{krot_fid}),
is calculated by recursive calls to a fast-Fourier transform routine.
Thanks to this algorithm, the matrix-vector multiplication
$U_{0,\delta K} \psi_0$ requires $O(N \ln N)$ operations instead
of $O(N^2)$, and thus 
allows to deal with very large system sizes. Our data to be presented
below correspond to system sizes of up to
$N \le  262144 = 2^{18}$ which still allowed to collect enough statistics
for the calculation of $\sigma^2(M)$.

We now present our numerical results.
Fig.~\ref{fig3} shows the distribution $P(M)$ of $M(t)$
in the regime $\Gamma < \lambda$ for different times. It is seen
that even though $P(M)$ is not normally distributed, it is still well
characterized by its variance. A calculation of $\sigma^2(M)$ is thus
meaningful.

\begin{figure}
\includegraphics[width=6.2cm]{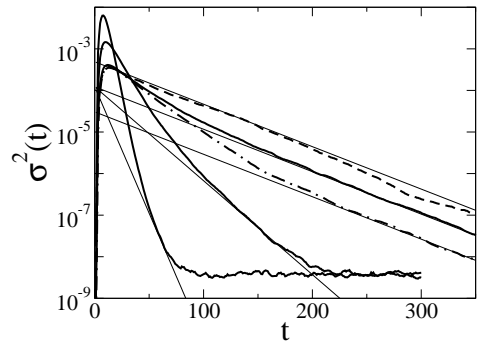}
\caption{\label{fig4} Variance
$\sigma^2(M)$ of the fidelity vs. $t$ for weak $\Gamma \ll \lambda$,
$N=16384$ and $10^{5}\cdot  \delta K = 5.9$, $8.9$ and $14.7$ (thick
solid lines), $N=4096$ and  $\delta K = 2.4 \cdot 10^{-4}$ (dashed line)
and $N=65536$ and $\delta K = 1.48 \cdot 10^{-5}$ (dotted-dashed line).
All data have $K_0=9.95$. The thin solid lines indicate the
decays $= 2 \hbar_{\rm eff} \exp[-\Gamma t]$, with $\Gamma = 0.024 
(\delta K \cdot N)^2$ (there is no adjustable free parameter).
The variance has been calculated from $10^3$ different
initial states $\psi_0$.\\[-1mm]}
\end{figure}

We next focus on
$\sigma^2$ in the golden rule regime with 
$\Gamma \ll \lambda$. Data are shown in Fig.~\ref{fig4}.
One sees that $\sigma^2(M)$ first rises up to a time
$t_c$, after which it decays. The maximal value 
$\sigma^2(t_c)$ in that regime increases with increasing perturbation,
i.e. increasing $\Gamma$. Beyond $t_c$, the decay of $\sigma^2$
is very well captured by Eq.~(\ref{sigma3g}), once enough
time has elapsed. This is due to the increase of $\sigma^2(t_c)$
above the self-averaging value $\propto \hbar_{\rm eff}$ as $\Gamma$ increases.
Once the influence of the peak
disappears, the decay of $\sigma^2(M)$ is very well captured by $\sigma^2_3$
given in Eq.~(\ref{sigma3g}), 
without any adjustable free parameter. Finally, at large times,
$\sigma^2(M)$ saturates at the value given in Eq.(\ref{eq:saturation}).

\begin{figure}
\includegraphics[width=6.2cm]{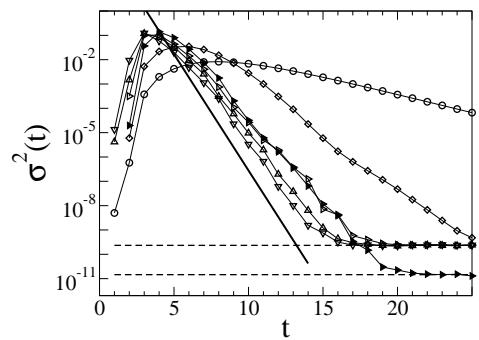}
\caption{\label{fig5}
Variance $\sigma^2(M)$ of the fidelity
vs. $t$ in the golden rule regime with $\Gamma \gtrsim \lambda$ for 
$N=65536$, $K_0=9.95$ and
$\delta K \in [3.9 \cdot 10^{-5},1.1 \cdot 10^{-3}]$ (open symbols),
and  $N=262144$, $K_0=9.95$, $\delta K=5.9 \cdot 10^{-5} $ (full triangles). 
The solid line is 
$\propto \exp[-2 \lambda_1 t]$, with an exponent $\lambda_1 = 1.1$, 
smaller than the Lyapunov exponent $\lambda=1.6$, because the fidelity
averages $\langle \exp[-\lambda t] \rangle$ (see text).
The two dashed lines give $\hbar_{\rm eff}^2=N^{-2}$.
In all cases, the variance has been calculated from $10^3$ different
initial states $\psi_0$. \\[-1mm]}
\end{figure}

\begin{figure}
\includegraphics[width=6.2cm]{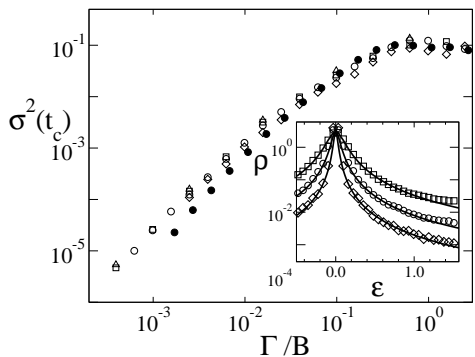}
\caption{\label{fig6}
Maximal variance $\sigma^2(t_c)$ as a function of 
$\Gamma/B$, for $K_0=10.45$, $N=4096$, 16384, 65536 and 262144
(empty symbols) and $K_0=50.45$, $N=16384$ (full circles). 
The variance has been calculated from $10^3$ different
initial states $\psi_0$. Inset: local spectral density of
states $\rho(\epsilon)$ 
of eigenstates of an unperturbed kicked rotator
with $K_0=12.56$ over the eigenstates of a perturbed
kicked rotator with $K=K_0+\delta K$, $\delta K=5 \cdot 10^{-3}$. System
sizes are $N=250$ (diamonds), $N=500$ (circles) and $N=1000$
(squares). The solid lines are Lorentzian
with widths $\Gamma \approx 0.0125$, $0.05$ and $0.0124$ in agreement
with the formula $\Gamma = 0.024 \; (\delta K \cdot N)^2$.
\\[-1mm]}
\end{figure}

As $\delta K$ increases, so does $\Gamma$ and $\sigma^2(M)$ decays faster
and faster to its saturation value until 
$\Gamma \gtrsim \lambda$. Once $\Gamma$ 
starts to exceed $\lambda$, the decay saturates at $\exp(-2 \lambda t)$. 
This is shown in Fig.~\ref{fig5}, which corroborates the Lyapunov decay
of $\sigma^2(M)$ predicted by Eqs.~(\ref{sigma1lambda}). Note that
in Fig.~\ref{fig5}, the decay exponent differs from the Lyapunov exponent 
$\lambda=\ln[K/2]$ due to the fact 
that the fidelity averages $\langle C_s \rangle \propto \langle 
\exp[-\lambda t] \rangle \ne \exp[-\langle \lambda \rangle t]$ over
finite-time fluctuations of the Lyapunov exponent 
\cite{silvestrov}. At long times, $\sigma^2(M)$ saturates at the 
ergodic value $\sigma^2(M,t \rightarrow \infty) = \hbar_{\rm eff}^2$,
as predicted. Finally, it is seen in both Figs.~\ref{fig4} and \ref{fig5}
that $t_c$ decreases as the perturbation is
cranked up. Moreover, there is no $N$-dependence of $\sigma^2(t_c)$ at fixed
$\Gamma$. These two facts are at least in qualitative, if not quantitative,
agreement with Eq.~(\ref{tc}). 

The behavior of $\sigma^2(t_c)$ as a function of $\Gamma$ is finally shown
in Fig.~\ref{fig6}. First we show in the inset the behavior of 
the local spectral density 
\begin{eqnarray}\label{ldos}
\rho(\epsilon) &=& \sum_\alpha |\langle \phi_\beta^{(0)} | \phi_\alpha 
\rangle|^2 \delta(\epsilon-\epsilon_\alpha+\epsilon_\beta),
\end{eqnarray}
of eigenstates $\{\phi_\alpha^{(0)}\}$ (with
quasienergy eigenvalues $\epsilon_\alpha$) of $U_{0}$
over the eigenstates $\{\phi_\alpha\}$ (with
quasienergy eigenvalues $\epsilon_\alpha^{(0)}$) of $U_{\delta K}$.
As mentioned above, $\rho(\epsilon)$ has a Lorentzian shape with a width
given by $\Gamma \approx 0.024 (\delta K \cdot N)^2$. Having extracted the
$N-$ and $\delta K-$dependence of $\Gamma$, we next plot
in the main part of Fig.~\ref{fig6} 
the maximum $\sigma^2(t_c)$ of the fidelity variance as a function 
of the rescaled width $\Gamma/B$ of $\rho(\epsilon)$.
As anticipated, $\sigma^2(t_c)$
first increases with $\Gamma$ until it saturates at a
value $\gtrsim 0.1$, independently on $\hbar_{\rm eff}$,
$\Gamma$ or $\lambda$, once $\Gamma \approx B$. These data confirm
Eq.~(\ref{sigmatc}) and the accompanying reasoning. Note that once
$\Gamma$ exceeds the bandwidth $B$, $\rho(\epsilon)$ is no longer
Lorentzian, and the decay of both $M(t)$ and $\sigma^2(M)$ is no
longer exponential \cite{Jac01}.

In conclusion we have applied both a semiclassical and a RMT approach to
calculate the variance $\sigma^2(M)$ 
of the fidelity $M(t)$ of Eq.~(\ref{fidelity}).
We found that $\sigma^2(M)$ exhibits a nonmonotonous behavior with time, first
increasing algebraically, before decaying exponentially at larger times.
The maximum value of $\sigma^2(M)$ 
is characterized by a non-self-averaging behavior when the perturbation
becomes sizable against the system's Lyapunov exponent.

This work was supported by the Swiss National Science Foundation. We 
thank J.-P. Eckmann and P. Wittwer for discussions on structural stability,
and \.I. Adagideli for discussions at the early stage
of this project.

\end{document}